\documentclass[twocolumn]{aastex631}

\usepackage{xspace}
\usepackage{subfigure}
\pdfoutput=1 %for arXiv submission
\usepackage{amsmath}
\usepackage[T1]{fontenc}
\usepackage{url}
\usepackage{graphicx}
\usepackage{lipsum} % just for the example
\usepackage{float}
\usepackage{bm}

\begin{document}
\setlength{\abovedisplayskip}{0pt}
\setlength{\belowdisplayskip}{2pt}

\title{Abundance Measurements of the Metal-poor M subdwarf LHS 174 Using High-resolution Optical Spectroscopy}

\author[0000-0001-5541-6087]{Neda Hejazi}
\affil{Department of Physics and Astronomy, University of Kansas, Lawrence,  KS 66045, USA}
\affil{Department of Physics and Astronomy, Georgia State University, Atlanta, GA 30303, USA}
\email{nhejazi@ku.edu}

\author[0000-0002-2437-2947]{S{\'e}bastien L{\'e}pine}
\affil{Department of Physics and Astronomy, Georgia State University, Atlanta, GA 30303, USA}

\author[0000-0001-5344-8069]{Thomas Nordlander}
\affil{Theoretical Astrophysics, Department of Physics and Astronomy, Uppsala University, Box 516, 751 20 Uppsala, Sweden}

\author[0000-0003-0193-2187]{Wei-Chun Jao}
\affil{Department of Physics and Astronomy, Georgia State University, Atlanta, GA 30303, USA}

\author[0000-0002-1221-5346]{David R. Coria}
\affil{Department of Physics and Astronomy, University of Kansas, Lawrence,  KS 66045, USA}

\author[0000-0002-9903-9911]{Kathryn V. Lester}
\affil{Department of Astronomy, Mount Holyoke College, South Hadley, MA 01075, USA}

\begin{abstract}
Metal-poor M subdwarfs are among the oldest stellar populations and carry valuable information about the chemical enrichment history of the Milky Way. The measurements of chemical abundances of these stars therefore provide essential insights into the  nucleosynthesis in the early stages of the Galaxy's formation.  We present the detailed spectroscopic analysis of a nearby metal-poor M subdwarf, LHS 174 from its high-resolution optical spectrum, and apply our previously developed spectral fitting code, \texttt{AutoSpecFit}, to measure the abundances of five elements:[O/H]=$-$0.519$\pm$0.081, [Ca/H]=$-$0.753$\pm$0.177, [Ti/H]=$-$0.711$\pm$0.144, [V/H]=$-$1.026$\pm$0.077, and [Fe/H]=$-$1.170$\pm$0.135. We  compare the abundances of O, Ti, and Fe derived from this work and those from previous studies and demonstrate the observed data is clearly better matched with the synthetic model generated based on our abundances than those from the other analyses.  The accuracy of inferred stellar abundances strongly depends on the accuracy of physical parameters,  which motivates us to develop a reliable technique to determine the parameters of low-mass M dwarfs more accurately ever than before and infer abundances with  smaller uncertainties. 

\end{abstract}


\section{Introduction}\label{sec:intro}
The stellar populations of the Milky Way are predominantly composed of low-mass M dwarfs, which constitute about $\sim$75{\%} of the total star count (\citealt{Henry_Jao_2024} and references therein). M dwarfs have main-sequence lifetimes much longer than the age of the Universe, which means that their surface abundances have not been affected by the mixing (dredge-up) processes that alter the surface composition of evolved giant stars. In addition, atomic diffusion \citep{Michaud1984} has a negligible effect on the atmospheric composition of M dwarfs (with deeper convection zones), as compared to more massive F, G, and K dwarf (with shallower convection zones) \citep{Gao2018, Deal2020, Moedas2022, Nordlander2024}. The chemical abundances of M dwarfs thus reflect the chemical composition of their progenitor molecular clouds, which makes them ideal tracers of the Galaxy's chemical enrichment history. 

Although individual M dwarfs maintain a steady chemistry over time, given their long lifespans, they have likely experienced gradual, long-term orbital migration within the Galaxy. Stellar orbits are, in general,  subject to perturbations due to the gravitational interaction with other stars or surrounding gas and dust as well as  the Galactic phenomena caused by spiral arms, giant molecular clouds, or other large-scale structures such as the tidal interaction of the Milky Way with a satellite galaxy. The orbital migration of stars may change the stellar distributions in various regions and affect the overall chemical structure of the Galaxy. In particular, the radial migration of stars in the disk can play an important role in the metallicity-age gradient across different radii \citep[e.g.][]{Haywood2008,Schonrich_Binney2009}. The study of long-lived M dwarfs in different stellar populations can therefore elucidate their migration history and  provide crucial clues on Galactic chemo-dynamical evolution.

 M dwarfs are primary targets for transit and radial velocity surveys of exoplanets, and are likely to host at least one orbiting planet \citep{Dressing2013, Dressing2015}. Given their ubiquity and stable chemical pattern over time,  M dwarfs can provide excellent sites to study star-planet chemical connections and probe the formation of planetary systems. However,  only a handful of small M-dwarf samples have reported abundances using careful high-resolution spectroscopy through spectral synthesis \citep{Souto2017,Souto2022,Abia2020,Shan2021,Hejazi2024,Tabernero2024}.  Most  studies relevant to the chemical properties of  stellar populations and star-planet compositional links have utilized the abundances of hotter F, G, and K dwarfs. The extension of such studies to numerous  M dwarfs would reveal whether previously inferred trends have any dependence on stellar mass and if these trends are also valid in the low-mass regime. It is then of critical importance to develop reliable techniques to measure the abundances of M dwarfs.

The abundance analysis of M dwarfs has proven to be  challenging. Owing to their intrinsic faintness,  the acquisition of high signal-to-noise ratio (SNR), high-resolution spectra requires the use of large telescopes and long exposure times, which poses a costly observing time investment. Archival high-resolution M-dwarf spectra are therefore limited to small samples of bright, nearby stars. Even with access to high-resolution, high signal-to-noise M dwarf spectra, there are other challenges involved in constraining the chemical abundance of M dwarfs. For example, the spectra of M dwarfs are dominated by millions of molecular lines, mostly governed by TiO bands in the optical and H$_{2}$O bands in the near-infrared (NIR), which are blended with many crucial atomic lines. In addition, this highly dense forest of molecular bands substantially obscures the smooth appearance of the true spectral continuum, making its identification  difficult. Instead, we must work with a pseudo-continuum whose dependence on physical stellar parameters and chemical abundances is not the same as that of the true continuum. This demands a more meticulous treatment for spectroscopic analyses of M dwarfs compared to F, G, and K dwarfs that present a well-defined continuum in their spectra.

Low-metallicity M dwarfs, which are kinematically and chemically associated with the Galactic halo, are much less abundant than more metal-rich counterparts (making up $\sim$0.25$\%$ of the total Galactic stellar population, \citealt{Reid_Hawley_2005}), and are, on average, farther away from the Sun and dimmer than disk stars. Consequently, only several small samples of metal-poor M dwarfs have been analyzed using high-resolution spectroscopy \citep{Woolf2005,Woolf2006,Woolf2009,Woolf2020,Schmidth2009,Rajpurohit2014}, leaving the detailed properties of these stars largely unconstrained.  On the other hand, larger samples have been studied using low-resolution spectroscopy, where metal-poor M dwarfs can be classified through a metallicity class sequence as (moderately metal-poor)  M subdwarfs (sdM), extreme M subdwarfs (esdM), and ultra M subdwarfs (usdM) based on the relative strength of CaH bands with respect to the strength of TiO bands (\citealt{Gizis1997,Lepine2007,Hejazi2020,Hejazi2022} and references therein).

In this study, we perform an in-depth, high-resolution analysis of a nearby (located at $\sim$27 pc from the Sun) and relatively bright (G=12.035), metal-poor M-type dwarf, LHS 174. The fundamental parameters, i.e., effective temperature (T$_{\rm eff}$), surface gravity (log($g$)), overall metallicity ([M/H]), and microturbulence ($\xi$), have previously been determined for this star using high-resolution (R$\sim$33000) optical spectroscopy (\citealt{Woolf2005}, hereafter WW05). The synthetic model (generated using the method outlined in Section \ref{sec:spectral_synthesis}) based on the parameter values inferred from WW05 is in good agreement with the star's observed spectrum over many spectral regions (Sections \ref{sec:phot_class_param} and \ref{sec:abundance_measurements}). WW05 also measured the titanium and iron abundances of LHS 174 along with the above physical parameters via an iterative procedure. \cite{Schmidth2009} (hereafter, Sch09) then measured the oxygen abundance of the star using the physical parameters and the abundances of Ti and Fe from WW05 by employing molecular TiO lines (Sections \ref{sec:abundance_measurements} and \ref{sec:discussion}). It should be noted that \cite{Kesseli2019} have also reported the effective temperature and metallicity values for LHS 174 using its medium-resolution, optical spectrum. While the effective temperature from \cite{Kesseli2019} is in agreement with that of WW05,  there is a significant difference in metallicity between these two studies (Table \ref{tab:properties}). We measure the abundances of the three elements O, Ti, and Fe using  our newly developed pipeline, \texttt{AutoSpecFit} \citep{Hejazi2024}, and compare them with those from the previously used methodologies. We also infer the abundances of two additional elements, Ca and V, for this star using our method. In this pilot study, we apply \texttt{AutoSpecFit}, which was originally developed using high-resolution NIR spectra, to the high-resolution optical spectrum of LHS 174, showing the pipeline's capability to implement line analyses over various wavelengths regimes.   

This paper is organized as follows. We briefly describe our observations and data reduction process in Section \ref{sec:obs}. The classification and physical parameters of LHS 174 from previous studies are reported in Section \ref{sec:phot_class_param}. The spectral synthesis process, selection of atomic and molecular line lists, and abundance formulations are presented in Section \ref{sec:spectral_synthesis}. The normalization and line selection procedures are detailed in Section  \ref{sec:line_selection}.  Section \ref{sec:abundance_measurements} outlines our methods used to measure elemental abundances and uncertainties for O, Ca, Ti, V, and Fe. Finally, we discuss our resulting abundances and compare them with the values from other studies in Section \ref{sec:discussion} and summarize the present study in Section \ref{sec:summary}.

\section{Observations}\label{sec:obs}
The optical spectrum of the target  was observed using the ARC Echelle spectrograph  (ARCES) of the Apache Point Observatory (APO) 3.5-m telescope on September 4th, 2017 (PI: Neda Hejazi and Sebastien Lepine). ARCES is a high-resolution, cross-dispersed spectrograph that provides optical spectra with wavelength coverage from 3200 to 10000 {\AA} at a resolving power of R$\sim$33000 (9 km s$^{-1}$) with the 1.6" slit. The efficiency of the entire telescope together with the spectrograph system is more than 2.2$\%$ at 647 nm and the spectrograph has an efficiency between 2$\%$ and 8$\%$ at 630 nm.

For the target,  six science frames, each with an exposure time of 600 s, and an average airmass of  1.03 were taken. For calibration, ten bias frames, three arc images (using a ThAr lamp), five blue and ten red flat field images\footnote{We took twice as many red flat field  as blue flat field frames  because M dwarfs have more flux in longer wavelengths than shorter ones.} were taken. We follow the steps described in the ARCES manual, provided by Karen Kinemuchi\footnote{\url{http://astronomy.nmsu.edu/apo-wiki/lib/exe/fetch.php?media=wiki:arces:kinemuchi_arces_cookbook.pdf}}, to reduce our data using standard \texttt{IRAF} routines. We perform cosmic ray removal,  bias subtraction, bad pixel mask correction, extraction and normalization of the master flat field spectrum, and extraction of the arc calibration spectra. We then extract the 1-D target spectra and divide them by the normalized master flat field.  Finally, we assign the corresponding arcs to the flat-fielded target spectra, which are then wavelength calibrated through the dispersion correction procedure.

We perform a polynomial fit to the spectrum of each echelle order, which approximately represents an empirically determined blaze function for that order. To remove the blaze,  we divide the spectrum  by the fit, which increases the accuracy of the subsequent continuum/pseudo-continuum normalization (Section \ref{sec:line_selection}) required for analysis. The target spectrum has an average SNR of 60, allowing for abundance measurements of some key elements.

\section{Classification and Physical Parameters of LHS 174}\label{sec:phot_class_param}
The metal-poor LHS 174 (PM J03307+3401) is a nearby, high-proper motion star, which is bright enough for high-resolution  spectroscopic analyses.  The astrometric, photometric, classification and physical parameters of this star are listed in Table \ref{tab:properties}. \cite{Gizis1997} classified LHS 174 by combining metallicity class and spectral type as sdM0.5. \cite{Kesseli2019} then reported a combined classification of sdM0.0 for this star. More recently, \cite{Hejazi2020}  classified the target as esdM1.0 using a template-fit method through  a set of empirically assembled M-dwarf classification templates based on the measurements of the TiO and CaH molecular bands. \cite{Zhong2015} presented a finer metallicity classification scale by splitting  each class dM (metal-rich M dwarf), sdM, esdM, and usdM into three subclasses, labeled by ``r'' for the metal-rich, ``s'' for the standard, and ``p'' for the metal-poor subclass (for example, sdM$_{\rm r}$, sdM$_{\rm s}$, sdM$_{\rm p}$), resulting 12 metal subclasses in total.  \cite{Hejazi2020} also employed this classification scheme, though simply numbered the 12 subclasses from 1 for the most metal-rich M dwarfs (dMr) to 12 for the most metal-poor ultra M subdwarfs (usdMp), and assigned a metal subclass of 7 (esdM$_{\rm r}$) to LHS 174.

The Galactic velocity components relative to the Galactic center, i.e.,  U (positive when pointing towards the Galactic center), V (positive when pointing in the direction of the Galactic disk orbital motion), and W (positive when pointing towards the north Galactic pole), of LHS 174\footnote{The Galactic velocity components of our target were calculated using its astrometry, i.e., coordinates (RA and DEC), proper motions in RA and DEC (PM$_{\rm RA}$ and PM$_{\rm DEC}$), distance (D) or parallax ($\pi$), and radial velocity (RV), as listed in Table \ref{tab:properties}. We employed the relevant python code as part of astrolibpy (https://github.com/segasai/astrolibpy), where the components can be corrected for the solar motion relative to the local standard of rest (LSR). An additional correction was then made for the motion of the LSR relative to the Galactic center [U,V,W]=[0,+220,0] km s$^{-1}$.}, [U,V,W]=[$-$177.42,$-$21.15,$-$32.47] km s$^{-1}$, indicate its association with the Halo.  The disk stars generally tend to move in nearly circular orbits around the Galactic center with V components around +220 km s$^{-1}$ and relatively low absolute values of U and W components, as opposed to the random motion of halo stars that move around the Galactic center in different directions (\citealt{Hejazi2022} and references therein). The star's relatively high absolute value of U component and negative value of V component (i.e., in the opposite direction of the disk orbital motion),  along with its low metallicity as compared to typical higher-metallicity  disk stars, rule out its disk membership.

 WW05 inferred the physical parameters of LHS 174 using its high-resolution, optical APO/ARCES spectrum. Since WW05 measured the abundances of only two elements (Ti and Fe), our main goal of re-observing the star was to obtain the spectrum with higher SNR and analyze more elements. Unfortunately, the poor weather condition (with a substantial cloud coverage) during a significant part of our observation did not allow us to fully achieve this goal. Nevertheless, we have been able to measure two more elements, i.e., Ca and V, using their corresponding atomic lines. WW05 derived the  T$_{\rm eff}$, log($g$), and [M/H] of the star using an iterative process along with abundances of Fe and  Ti (obtained by measuring their equivalent width). The microturbulence ($\xi$) parameter was determined by requiring that there should be no slope in Ti abundance as a function of equivalent width. We generate the synthetic spectrum  associated with the determined parameters by WW05 as shown in Table  \ref{tab:properties} (Section \ref{sec:spectral_synthesis}). We find that the synthetic model exhibits an overall good consistency with the observed flux over many spectral lines, which indicates that these parameter values reliably present the atmospheric properties of the star.

However,  there is no uncertainty associated with the [M/H] and $\xi$ values from WW05.  We calculate the metallicity of LHS 174 by applying the photometric M-dwarf metallicity relation from \cite{Duque-Arribas2023} using the absolute \textit{G} magnitude and color \textit{$G_{BP}-G_{RP}$}, [M/H]$_{\rm Phot}$=$-$1.09 dex,  which is different from the metallicity reported in WW05, i.e., [M/H]$_{\rm WW05}$=$-$0.95 dex, by 0.14 dex. We therefore assume an uncertainty of 0.14 dex for the star's metallicity, which is larger than the abundance errors in  WW05. We also assume an uncertainty of 0.10 km s$^{-1}$ for the microturbulence parameter. The assumed uncertainties of [M/H] and $\xi$ are nearly the same or even larger than the typical errors from previous high-resolution analyses of M dwarfs \citep{Lindgren2016,Lindgren2017,Souto2017,Souto2020,Marfil2021,Hejazi2023,Hejazi2024}.

We further modify the uncertainties of T$_{\rm eff}$ and log($g$) reported in WW05 by deriving these parameters again from the current empirical photometric relations. We compute the effective temperature of the star by employing the photometric relation from \cite{Mann2015} using the colors \textit{$G_{BP}-G_{RP}$} and \textit{J-H}, (T$_{\rm eff}$)$_{\rm Phot}$=3855 K, which is different from the value presented in WW05, i.e., (T$_{\rm eff}$)$_{\rm WW05}$=3790 K, by 65 K. The T$_{\rm eff}$ error (20 K) derived by WW05 is thus  too small and we increase this error to 65 K.  On the other hand, the uncertainty of surface gravity from WW05 (0.31 dex) is unusually large  compared to the average error of log($g$) from other M-dwarf studies as cited above. This is mainly due to the significantly large uncertainty of the parallax ($\pi$=22.6$\pm$7.4 mas, which was available at the time of the WW05's study) that, along with \textit{K} magnitude and bolometric correction BC$_{K}$, was  used to calculate the bolometric magnitude (M$_{bol}$). The resulting M$_{bol}$, together with the star's mass and T$_{\rm eff}$, was then employed to derive the star's surface gravity. We derive the radius and mass of the star by utilizing the photometric relations from  \cite{Mann2015} and \cite{Mann2019}, respectively, using the absolute \textit{K} magnitude (\textit{M$_{K}$}) obtained by the  \textit{K} magnitude and the highly accurate Gaia-DR3 parallax ($\pi$=36.654$\pm$0.034 mas, \citealt{GaiaDR3}). The inferred radius and mass are used to determine the surface gravity, log($g$)$_{\rm Phot}$=4.90, which is different from the value determined by WW05, i.e., log($g$)=4.78, by 0.12 dex, and quite similar to the typical uncertainties reported the above-mentioned M-dwarf studies. We thus assume an error of 0.12 dex for surface gravity to be used in this work. All adopted parameter uncertainties are listed in Table \ref{tab:properties}.

We apply the rule for $\alpha$ enhancement,  [$\alpha$/Fe], described in \cite{Gustafsson2008}, where $\alpha$ elements have been scaled as follows:  [$\alpha$/Fe]=+0.4 for  $-$5.0$\leq$[M/H]$\leq$$-$1.0, [$\alpha$/Fe]=$-$0.4$\times$[M/H] for $-$1.0$\leq$[M/H]$\leq$0.0, and [$\alpha$/Fe]=0.0 for [M/H]$\geq$0.0. For [M/H]=$-$0.95, the $\alpha$ enhancement is  [$\alpha$/Fe]=+0.38, which is used an input for the abundance measurements. We adjust [$\alpha$/Fe] when [M/H] is deviated by its error for measuring the systematic errors of abundances (Section \ref{sec:abundance_errors}).

\section{Spectral Synthesis, Abundance Formulation, And Line Data}\label{sec:spectral_synthesis}
For our abundance analysis, we synthesize model spectra using \texttt{Turbospectrum} \citep{AlvarezPlez1998, Plez2012}, a one-dimensional (1D), local thermodynamic equilibrium (LTE) radiative transfer code, together with 1D, hydrostatic, and plane-parallel MARCS model atmospheres \citep{Gustafsson2008} and a selected set of atomic and molecular line lists. We employ the solar abundances reported in \cite{Asplund2021}, which are updated abundances with respect to the values from \cite{Grevesse2007}  used in our previous studies (\citealt{Hejazi2023,Hejazi2024,Hejazi2025}, hereafter, H23, H24, and H25, respectively). In this analysis, we use  ``continuum-normalized'' synthetic spectra (hereafter, synthetic or model spectra, for simplicity) computed by \texttt{Turbospectrum}, where line depths are measured from unity.

We utilize the interpolation routine developed by Thomas Masseron\footnote{https://marcs.astro.uu.se/software.php} to interpolate the MARCS model associated with the star's physical parameters from WW05 listed in Table  \ref{tab:properties}.  Synthetic spectra are generated using the interpolated model while abundances are customized by varying relative abundance ``ABUND(X)'' for element X, which can be converted to the ``absolute abundance'' A(X) by 

\begin{equation}\label{equ:absolute_abundance}
\begin{split}
 {\rm A(X)} &=  {\rm log(N_{ X}/N_{\rm H})+12} \\ 
 &= {\rm ABUND(X) + A(X)_{\sun} +  [M/H]}
\end{split}
\end{equation}

\noindent
where  N$_{\rm X}$ indicates the number density of element X, N$_{\rm H}$ indicates the number density of hydrogen, A(X)$_{\sun}$ is the solar absolute abundance of element X, and [M/H] is the star's overall metallicity.
Similarly, if X is an $\alpha$ element:

\
\begin{equation}\label{equ:abundance_with_alpha}
{\rm A(X)}= {\rm ABUND(X) + A(X)_{\sun} +  [M/H] + [\alpha/Fe]}
\end{equation}

\noindent
where [$\alpha$/Fe] is the star's $\alpha$-element enhancement\footnote{In our previous studies (H23, H24, and H25), the $\alpha$ enhancement [$\alpha$/Fe] of the planet-host cool dwarfs with solar or super-solar metallicity was assumed to be zero, and was removed in Eq. \ref{equ:abundance_with_alpha}.}. The ``abundance''  [X/H] is computed relative to the solar value using the equation:

\begin{equation}\label{equ:abundance}
\begin{split}
 {\rm [X/H]} &=  {\rm log({N_{X}}/{N_{H}}) -log({N_{X}}/{N_{H}})_{\sun}} \\ 
 &= {\rm A(X)-A(X)_{\sun}}\\
 &= {\rm ABUND(X) + [M/H]}
\end{split}
\end{equation}

\noindent
or
\begin{equation}\label{equ:abundance_with_alpha}
{\rm [X/H]}= {\rm ABUND(X) + [M/H] + [\alpha/Fe]}
\end{equation}
\
\noindent
if X is an $\alpha$-element.

We employ the atomic line list from the Vienna Atomic Line Database (VALD3\footnote{https://vald.astro.uu.se/}, \citealt{Ryabchikova2015}), where hyperfine structure (HFS) data have recently been added \citep{Pakhomov2017, Pakhomov2019}. We have found slight differences between the atomic line data from VALD3 and \texttt{Linemake}\footnote{The atomic datasets presented in \texttt{Linemake}  primarily  originate from the publications of the atomic physics group in the University of Wisconsin (for example, \citealt{Lawler2009,Lawler2013}, among others), and have been updated over time.} \citep{Placco2021} for some of our analyzed spectral lines, and we have accordingly replaced the potential energy of the lower level and log($gf$), where g is the statistical weight of the lower level and f is the oscillator strength of the transition, reported in VALD3 by those values reported in \texttt{Linemake}  (Table \ref{tab:line_data}). 

We also use the most recent molecular TiO line list (the ExoMol TOTO) from \cite{McKemmish2019} where a new line database for the main isotopologues of TiO, i.e., $^{46}$Ti$^{16}$O, $^{47}$Ti$^{16}$O, $^{48}$Ti$^{16}$O, $^{49}$Ti$^{16}$O, $^{50}$Ti$^{16}$O has been presented. The ExoMol TOTO line list includes all
dipole-allowed transitions between 13 low-lying electronic states (X$^{3}\Delta$, a$^{1}\Delta$, d$^{1}\Sigma^{+}$, E$^{3}\Pi$, A$^{3}\Phi$, B$^{3}\Pi$, C$^{3}\Delta$, b$^{1}\Pi$, c$^{1}\Phi$, f$^{1}\Delta$, and e$^{1}\Sigma^{+}$). The rovibronic line positrons were obtained by employing potential energy curves (as simple Morse oscillators\footnote{A Morse Oscillator is a specific kind of bond stretch oscillator commonly used to simulate the anharmonic stretching vibrations  of polyatomic molecules in quantum mechanics.})  with constant diagonal and off-diagonal spin-orbit and other coupling terms fitted to known empirical energy levels, or to computed ab initio curves if the experimental data were unavailable. The ExoMol TOTO line lists are suitable  for effective temperatures below 5000 K, including  nearly 60 million transitions for the major TiO isotopologues. However, due to the dissociation of TiO molecules in  stellar atmospheres with T$_{\rm eff}$>4200 K, TiO line data are not needed to analyze these stars.

\begin{figure*}[h]
    \centering
    \includegraphics[width=1\linewidth]{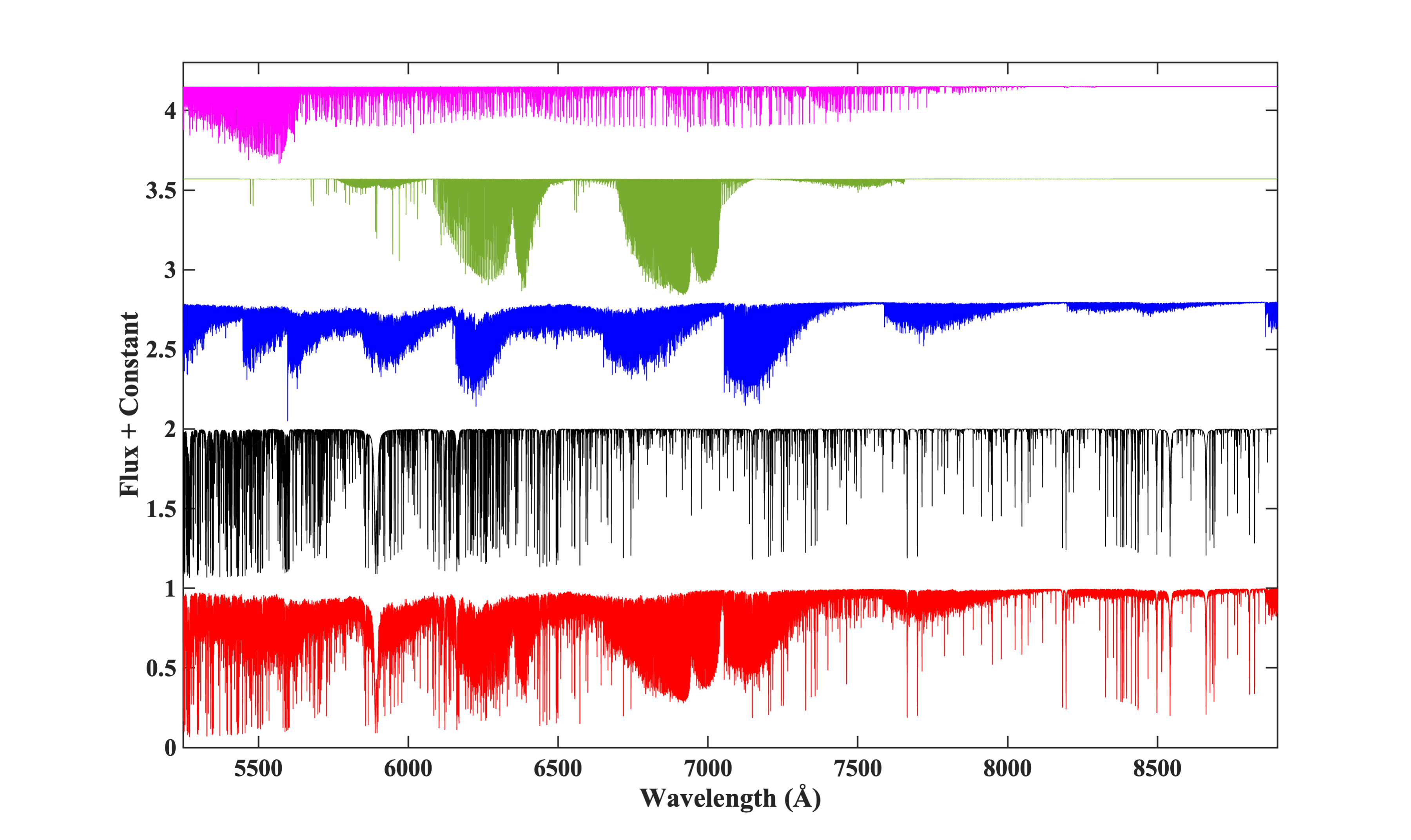}
    \caption{Synthetic models associated with the star's parameters and default abundances calculated from Equations \ref{equ:absolute_abundance} or \ref{equ:abundance} assuming ABUND(X)=0 for all elements X when including: only MgH molecular lines (magenta); only CaH molecular lines (green); only TiO molecular lines (blue); only atomic lines (black); and all atomic and molecular lines (red).}
\label{fig:models}
\end{figure*}

\cite{Pavlenko2020} have showed that  the ExoMol TOTO line list describes the fine details in the line position and intensity of M-dwarf spectra more accurately than other TiO line lists.  To make our spectral synthesis computationally efficient,  we significantly reduce the extremely large number of lines in the ExoMol TOTO line list by choosing stronger lines using the cutoff parameter ``a'' \citep{Pavlenko2020}:

\begin{equation}\label{equ:line_cutoff}
{\rm a}={\rm (gf) \times exp(-E_{low}/kT)} > 10^{-6}\\   
\end{equation}

\noindent
where E$_{\rm low}$ is the energy of the lower state, k is the Boltzmann constant, and T=3500 K that is the typical effective temperature of M dwarfs. We also include the line lists of other molecular bands such as VO and CaH (computed by Bertrand Plez, 1998), CrH \citep{Burrows2002}, MgH (\citealt{Skory2003} for $^{24}$MgH and \citealt{Kurucz2011} for $^{25}$MgH and $^{26}$MgH), and several more hydrides and oxides. Figure \ref{fig:models} shows the generated synthetic models associated with the star's fundamental parameters and default abundances inferred from Equations \ref{equ:absolute_abundance} or \ref{equ:abundance} assuming ABUND(X)=0 for all elements X, when only MgH molecular lines (magenta), only CaH molecular lines (green), only TiO molecular lines (blue), only atomic lines (black), and all atomic and molecular lines (red) are incorporated.

\section{Normalization and Line Selection}\label{sec:line_selection}
We perform a careful visual inspection over the entire optical spectrum (3200 to 10000 {\AA}) of LHS 174.  We first remove the spectral ranges that include artifacts, bad pixels, strong noise, and telluric absorption lines due to the Earth's atmosphere.  We then select the spectral lines that are strong enough to be identified from the dominant background molecular lines and are almost isolated from neighboring lines. To reduce the contribution of blending TiO molecular lines, we exclude those atomic lines that overlap with TiO lines having depths greater than 20$\%$ from unity.

The red spectrum in Figure \ref{fig:models} roughly presents the star's best-fit model (hereafter, Model$_{\rm Approx}$) and can be used to select the best lines for abundance measurements. We perform a trial-and-error examination for different wavelength intervals within the continuum, or for most cases, pseudo-continuum near each side of the lines, which must include at least one or two wavelength data points. These intervals are used in the normalization routine, where the observed spectrum is normalized relative to Model$_{\rm Approx}$ using a linear fitting (see H24 and H25 for details). The most appropriate continuum/pseudo-continuum ``normalizing intervals'' are selected when a good agreement between the normalized observed flux and Model$_{\rm Approx}$ across these ranges is achieved.

We carry out an additional inspection for the initially selected normalizing intervals around each  line  and vary the corresponding relative abundance ABUND(X) by $\Delta$ABUND(X)=$\pm$0.300 dex\footnote{This variation range is more than sufficient because the values of  ABUND(X) are typically in the range [$-$0.200,+0.200] dex in this analysis as well as our previous studies.}  and explore whether the data points in these intervals still show a consistency between the  normalized observed data and  synthetic model  or a considerable discrepancy is evident. On occasions, changing the abundances of specific elements such as oxygen or iron may cause a few weak lines to emerge inside the pseudo-continuum intervals that were apparently line-free regions before the abundance variation.  We re-select the normalizing intervals if the abundance variation makes an inconsistency between the normalized observed flux and synthetic model spectrum or/and gives rise to a significant  deformation of the pseudo-continuum. It is important to note that in some wavelength areas, there are a few (useful) spectral lines that are rather close to each other and can be normalized using the same normalizing intervals, usually in both sides of, and in some cases, also inside those areas. The final selected normalizing intervals are used in normalizing the star's observed spectrum for the following steps of our analysis.

 We  manually perform spectral fitting over the lines by varying their corresponding relative abundances from ABUND(X)=$-$0.300 dex to ABUND(X)=+0.300 dex in steps of 0.100 dex and comparing the normalized observed data with the model spectra.  Through visual investigation, we exclude those lines that cannot be fitted with any synthetic model in depth and/or shape. The evident discrepancies can be due to the deficiency in model atmospheres and/or line opacity datasets of low-mass M dwarfs.  Other factors such as  deviations from LTE may also be a major reason for differences between the observed and model spectra over some spectral lines \citep{Olander2021}. High noise levels and strong telluric absorption (and non-stellar) lines   distort the star's spectrum within some wavelength ranges that cannot be matched with model spectra.
 
 During running \texttt{AutoSpecFit} (Section \ref{sec:abundance_measurements}), we find a few  lines that do not result in a well-defined minimum  ${\chi}^{2}$ value and we remove these lines from the analysis. our final selected lines  include 28 lines associated with five elements: O (using TiO molecular bands, see Section \ref{sec:discussion} for details), Ca I (and also one Ca II), Ti I, V I, and Fe I, as shown in Table \ref{tab:line_data}. We search for hyperfine structure (HFS) states of these lines\footnote{The HFS splitting lines are not included in our original atomic line list; instead, the average of all HFS lines as a single line is presented.} using the VALD3 website through  the \texttt{Extract Stellar} option, and find that only the two vanadium lines are split into HFS splitting lines. We incorporate these HFS lines in our atomic line list, as shown in the second column of Table \ref{tab:line_data}. The log($gf$) values of the selected lines are listed in the third column of the table. In addition, for each analyzed line, we determine a fitting or ${\chi}^{2}$ window, where the ${\chi}^{2}$ minimization process is performed, as presented in the fourth column of the table.

\begin{deluxetable*}{llcl}\label{tab:properties}
%\tablenum{1}
\tablecaption{The astrometry, photometry, classification and physical parameters of LHS 174}
\tablewidth{0pt}
\tablehead{
Property's name & Description & \colhead{Value} & \colhead{Reference}   
}
\startdata
\textbf{Astrometry} &   &   & \\
 \hline
RA & Right ascension &  03 30 44.82  &   \cite{GaiaDR3} \\
DEC & Declination &  +34 01 07.19  &   \cite{GaiaDR3} \\
PM$_{\rm RA}$ (mas yr$^{-1}$)  & Proper motion in RA & 497.882 & \cite{GaiaDR3} \\
PM$_{\rm DEC}$ (mas yr$^{-1}$)  & Proper motion in DEC & $-$1499.311 & \cite{GaiaDR3} \\
$\pi$ (mas) & Parallax &  36.654 & \cite{GaiaDR3} \\
D (pc)  & Distance & 27.28 & \cite{GaiaDR3} \\
RV (km s$^{-1}$) & Radial velocity & $-$230.43 & \cite{GaiaDR3} \\
\hline
\textbf{Photometry} &  &    &  \\
  \hline
\textit{G} & Gaia G magnitude & 12.035 &  \cite{GaiaDR3} \\
\textit{G$_{BP}$} & Gaia BP magnitude & 12.980 &  \cite{GaiaDR3} \\
\textit{G$_{RP}$} & Gaia RP magnitude & 11.070 &  \cite{GaiaDR3} \\
\textit{J} & 2MASS J magnitude & 9.844 & \cite{Cutri2003} \\
\textit{H} & 2MASS H magnitude & 9.350 & \cite{Cutri2003} \\
\textit{K} & 2MASS K magnitude & 9.143 & \cite{Cutri2003} \\
\hline
\textbf{Classification} &   &   &  \\
\hline
MC+SpType & Metallicity class+Spectral type &  sdM0.5  & \cite{Gizis1997} \\
MC+SpType & Metallicity class+Spectral type &  sdM0.0  & {\cite{Kesseli2019}} \\
MC+SpType & Metallicity class+Spectral type &  esdM1.0 & \cite{Hejazi2020} \\
SubMC & Metallicity subclass & 7 (esdM$_{\rm r}$)  & \cite{Hejazi2020} \\
\hline
\textbf{Physical Parameters} &  &   &  \\
\hline
T$_{\rm eff}$ (K)  & Effective temperature &  3790$\pm$65* &  \cite{Woolf2005}   \\
T$_{\rm eff}$ (K)  & Effective temperature & 3800$\pm$100 & \cite{Kesseli2019} \\
$[$M/H$]$  &  Overall metallicity  & $-$0.95$\pm$0.14* & \cite{Woolf2005} \\
$[$M/H$]$  &  Overall metallicity  &  $-$0.63$\pm$0.30 & \cite{Kesseli2019} \\
log($g$)   &  Surface gravity & 4.78$\pm$0.12* & \cite{Woolf2005} \\
$\xi$ (km s$^{-1}$)  & Microturbulent Velocity & 1.00$\pm$0.10*  &  \cite{Woolf2005} \\
\enddata
\tablecomments{*There is no reported uncertainty for $[$M/H$]$ and $\xi$ in \cite{Woolf2005} and  an  error value  for each of these two parameters have been assigned. The original uncertainties of T$_{\rm eff}$ and log($g$) presented by \cite{Woolf2005} have been modified (Section \ref{sec:phot_class_param}).} 
\end{deluxetable*}

\begin{deluxetable*}{lcccc}\label{tab:line_data} 
%\tablenum{2}
\tablecaption{The selected lines used to measure the abundances of LHS 174}  
\tablewidth{0pt}
\tabletypesize{\scriptsize}
\tablehead{
\colhead{Species} &  Central wavelength ({\AA}) & log($gf$) & {$\chi^{2}$} window ({\AA})  & Comments}
\startdata
O (using TiO) & 7080.220  & [$-$5.238,0.440]$^{a}$ &  7079.98-7080.42  & The combination of 475 TiO lines$^{a}$    \\
O (using TiO)  & 7081.440  & [$-$5.002,0.550]$^{a}$ &  7081.24-7081.65  & The combination of 435 TiO lines$^{a}$  \\
\hline
Ca I$^{b}$   & 6102.723  & $-$0.820 & 6102.15-6103.25 &  Blended with a weak Ca I line: 6102.439   \\
Ca I$^{b}$   & 6122.217  & $-$0.340 & 6121.50-6123.00 &    \\
Ca I   & 6439.075 &   +0.390 & 6438.35-6439.93 &    \\
Ca I   & 6455.598  & $-$1.340 & 6455.27-6455.88 &    \\
Ca I   & 6462.567  & +0.262 & 6462.11-6463.03 &  Blended with a weak Fe I line: 6462.725  \\
Ca I   & 6471.662  & $-$0.686 & 6471.40-6471.95 &    \\
Ca I   &  6493.781  & $-$0.109 & 6493.40-6494.15 &    \\
Ca I   & 6499.650  & $-$0.818 & 6499.35-6499.90 &    \\
Ca I$^{b}$   & 6572.779  & $-$4.320 & 6572.40-6573.15 & Blended with a weak Cr I line: 6572.885   \\
Ca II$^{b}$  & 8498.020  & $-$1.360 & 8497.35-8498.70 &  Blended with four weak Ca II lines:\\
       &          &          &             &    8498.077, 8498.098,  8498.129, 8498.233            \\
\hline
Ti I   & 6554.223  & $-$1.150 & 6554.05-6554.45 &    \\
Ti I   & 6556.062  & $-$1.060  & 6555.85-6556.30 &    \\
Ti I   & 8377.861  & $-$1.590 & 8377.58-8378.25 &    \\
Ti I$^{b}$   & 8396.897  & $-$1.780 & 8396.60-8397.25 &    \\
Ti I$^{b}$   & 8412.357  & $-$1.480 & 8412.00-8412.75 &    \\
Ti I$^{b}$   & 8426.507  & $-$1.250 & 8426.15-8426.85 &    \\
Ti I   & 8682.983  & $-$1.790 & 8682.73-8683.25 &    \\
\hline
V I$^{c}$    &              &   & 6089.94-6090.45 & The combination of 18 HFS V I lines   \\
       &  6090.19386    & $-$0.700  &                &              \\
       &  6090.20104    & $-$0.841  &                &              \\ 
       &  6090.20736    & $-$1.005  &                &              \\ 
       &  6090.20766    & $-$1.540  &                &              \\ 
       &  6090.21283    & $-$1.203  &                &              \\ 
       &  6090.21292    & $-$1.344  &                &              \\ 
       &  6090.21729    & $-$1.290  &                &              \\ 
       &  6090.21743    & $-$1.458  &                &              \\ 
       &  6090.21953    & $-$2.654  &                &              \\ 
       &  6090.22079    & $-$1.312  &                &              \\ 
       &  6090.22117    & $-$1.846  &                &              \\ 
       &  6090.22285    & $-$2.244  &                &              \\ 
       &  6090.22342    & $-$1.403  &                &              \\ 
       &  6090.22516    & $-$1.591  &                &              \\ 
       &  6090.22526    & $-$2.022  &                &              \\ 
       &  6090.22678    & $-$1.897  &                &              \\ 
       &  6090.22717    & $-$1.876  &                &              \\ 
       &  6090.22742    & $-$1.846  &                &              \\ 
V I$^{c}$    &              &   & 6111.33-6111.97 & The combination of 4 HFS V lines \\
       &  6111.59249    & $-$1.701  &                &               \\
       &  6111.63213    & $-$1.224  &                &              \\ 
       &  6111.65616    & $-$1.224  &                &              \\ 
       &  6111.69580    & $-$1.370  &                &              \\ 
\hline
Fe I    & 5371.489  & $-$1.645 &  5371.15-5371.80 &    \\
Fe I$^{b}$    & 5434.523  & $-$2.130 & 5434.03-5434.87 &    \\
Fe I$^{b}$    & 6430.845  & $-$1.950 & 6430.60-6431.20 &    \\
Fe I$^{b}$    & 8047.617  & $-$4.660 & 8047.42-8047.85 &    \\
Fe I$^{b}$    & 8387.771  & $-$1.510 & 8387.25-8388.25 &    \\
Fe I$^{b}$    & 8688.623  & $-$1.200 & 8688.10-8689.15 &    \\
Fe I    & 8824.220  & $-$1.540 & 8823.75-8824.65 &    \\
\enddata
\tablecomments{$^{a}$These ranges and line numbers are associated with the region within the {$\chi^{2}$} window and only the molecular lines of the most abundant isotopologue, $^{48}$TiO, are accounted, though all the five main TiO isotopologue are used to generate our model spectra. $^{b}$The line data originate from \texttt{Linemake}. $^{c}$Our source VLAD atomic line list used in this study shows the wavelengths in three decimal digits. However, for better presentation, we show the HFS lines of vanadium in five decimal digits to distinguish the lines that are very close to each other.} 
\end{deluxetable*}

\section{Abundance Analysis}\label{sec:abundance_measurements}

\subsection{Elemental Abundance Measurements}
 To measure the elemental abundances of LHS 174, we apply the newly refined version of our automatic spectral fitting code, \texttt{AutoSpecFit} (H24)\footnote{We have recently modified the code slightly for the weighted average abundance of different lines corresponding to each analyzed element in each iteration of the procedure.}, which performs a series of line-by-line ${\chi}^{2}$ minimization processes in an iterative manner. The physical parameters (Section \ref{sec:phot_class_param}), the selected normalizing intervals, and the selected ${\chi}^{2}$ windows associated with the analyzed lines are employed as input data to run \texttt{AutoSpecFit}. The pipeline allows Turbospectrum to generate the required synthetic models  based on all abundances inferred from each iteration ``on the fly'';  these synthetic models are then used in the next iteration. Prior to computing ${\chi}^{2}$ values,  the synthetic spectra are smoothed at the observed spectral resolution using a Gaussian kernel and interpolated at wavelengths that shifted by the star's radial velocity. The code then carries out a line-by-line normalization of the observed spectrum with respect to all synthetic models examined in each iteration using the input normalizing intervals. The relative abundance of element X  varies from ABUND(X)=$-$0.400 dex to ABUND(X)=+0.400 dex in steps of 0.020 dex, and a polynomial fit is implemented to the resulting ${\chi}^{2}$ values as a function of ABUND(X) for each relevant line. A relative abundance of ABUND(X)$_{\rm Line}$ is assigned to each line, for which ${\chi}^{2}$ is minimized. The ABUND(X) value for element X is the weighted average of ABUND(X)$_{\rm Line}$ over all  spectral lines corresponding to element X and is optimized through consecutive iterations until the relative abundances of all analyzed elements converge to their final values simultaneously (see H24 for a comprehensive description of the method). It is important to note that, in general, there is a complex correlation between the abundances of different elements and simultaneously measuring elemental abundances during an iterative procedure is the best approach to consider this correlation.

Elemental abundances are highly degenerate with physical parameters, and the most reliable  method is to keep the parameters fixed when varying abundances. We will further add a routine to the future version of \texttt{AutoSpecFit} to  modify the parameters between the iterations of abundance inference. In our follow-up study, we investigate the degeneracy between physical parameters and find the spectral regions that are mostly sensitive to only one parameter. The use of these regions would separate the contribution of different parameters that are simultaneously varied in the spectral fitting and would significantly reduce the parameter degeneracy, leading to more accurate  stellar parameters.

\begin{deluxetable*}{l|c|c|c|c|c|c}[htp]\label{tab:results}
\tablecaption{The chemical abundances of the five analyzed elements and their sensitivity to the variation of  effective temperature, overall metallicity, surface gravity, and microturbulence as well as the  systematic errors $\sigma_{\rm{sys}}$, random errors $\sigma_{\rm{ran}}$, and total uncertainties $\sigma_{\rm{tot}}$ for LHS 174}  
\tablewidth{0pt}
\tabletypesize{\scriptsize}
\tablehead{{Species} & Description  & {O (using TiO)} & {Ca}  & {Ti} & {V} & {Fe}}
\startdata
N & Number of analyzed lines & 2 & 10 & 7 & 2 & 7\\
\hline
{[X/H]} & Abundance form this work &  $-$0.519$\pm$0.081 & $-$0.753$\pm$0.177  & $-$0.711$\pm$0.144 & $-$1.026$\pm$0.077 & $-$1.170$\pm$0.135  \\
\hline
{[X/H]} & Abundance from previous studies &  $-$0.37$\pm$0.17  & ... & $-$0.83$\pm$0.07  & ... & $-$1.11$\pm$0.05  \\
   & References  &  Sch09 & ... & WW05 & ... & WW05 \\
\hline
${\rm{\Delta}{\rm[X/H]}_{\rm{T,Neg}}}$  & Abundance change if {$\rm{{\Delta}T_{eff}}$}=$-$65 K & $-$0.049 & $-$0.128 & $-$0.101 & $-$0.056 & $-$0.072 \\
${\rm{\Delta}{\rm[X/H]}_{\rm{T,Pos}}}$  & Abundance change if {$\rm{{\Delta}T_{eff}}$}=+65 K & +0.058 & +0.144 & +0.114 & +0.064 & +0.079 \\
 $\overline{\rm{({\Delta}[X/H])_{T}}}$ & Average abundance change   & 0.054 & 0.136 & 0.108  & 0.060 & 0.076 \\
\hline
${\rm{\Delta}{\rm[X/H]}_{\rm{M,Neg}}}$  & Abundance change if {$\rm{{\Delta}[M/H]}$}=$-$0.14 dex & $-$0.045 & $-$0.056 & $-$0.070  & $-$0.047 & $-$0.087  \\
${\rm{\Delta}{\rm[X/H]}_{\rm{M,Pos}}}$  & Abundance change if {$\rm{{\Delta}[M/H]}$}=+0.14 dex & +0.050  & +0.068  & +0.066  & +0.047 & +0.084 \\
 $\overline{\rm{({\Delta}[X/H])_{M}}}$ & Average abundance change   & 0.048 & 0.062 & 0.068  & 0.047 & 0.086 \\
\hline
${\rm{\Delta}{\rm[X/H]}_{\rm{G,Neg}}}$  & Abundance change if {$\rm{{\Delta}}$log($g$)}=$-$0.12 dex & $-$0.036  & +0.104   & +0.056  & +0.000  & +0.067 \\
${\rm{\Delta}{\rm[X/H]}_{\rm{G,Pos}}}$  & Abundance change if {$\rm{{\Delta}}$log($g$)}=+0.12 dex & +0.036 & $-$0.078 & $-$0.041 & +0.006 & $-$0.052  \\
$\overline{\rm{({\Delta}[X/H])_{G}}}$ & Average abundance change  & 0.036 & 0.091  & 0.049 & 0.003 & 0.060  \\
\hline
${\rm{\Delta}{\rm[X/H]}_{{\xi,\rm{Neg}}}}$  & Abundance change if {$\rm{{\Delta}{\xi}}$}=$-$0.10 km s$^{-1}$ & $-$0.011 & +0.003  & +0.016 & +0.008 & +0.007 \\
${\rm{\Delta}{\rm[X/H]}_{{\xi,\rm{Pos}}}}$  & Abundance change if {$\rm{{\Delta}{\xi}}$}=+0.10 km s$^{-1}$ & +0.011 & $-$0.003  & $-$0.018 & $-$0.008 & $-$0.005  \\
$\overline{\rm{({\Delta}[X/H])_{\xi}}}$ & Average abundance change  & 0.011 & 0.003 & 0.017 & 0.008 & 0.006 \\
\hline
$\sigma_{\rm{sys}}$  & Systematic error & 0.081 & 0.175  & 0.138  & 0.077 & 0.130  \\
\hline
$\sigma_{\rm{ran}}=\rm{std}/\sqrt{N}$  & Random error & ... & 0.029   & 0.040  & ... & 0.037 \\
\hline
$\sigma_{\rm{tot}}$   & Total uncertainty & 0.081 & 0.177  & 0.144  & 0.077 & 0.135 \\
\enddata
\end{deluxetable*}

\begin{figure*}[htp]
    \centering
    \includegraphics[width=1\linewidth]{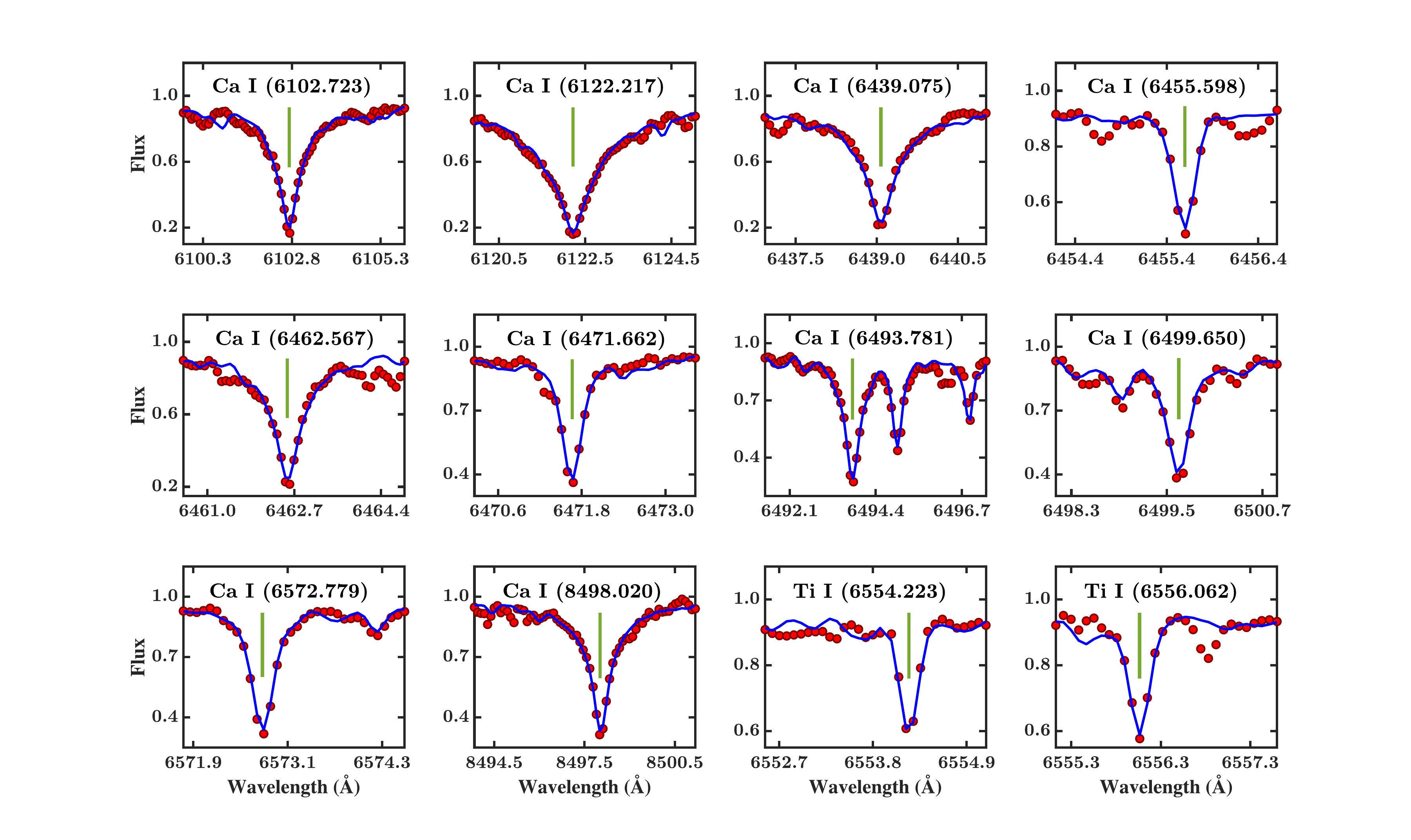}
    \caption{Comparison between the observed spectrum (red dots) and the best-fit model (blue lines) for the analyzed Ca I and Ti I lines of LHS 174. The observed flux is normalized to the best-fit model.}
\label{fig:bestfit_1}
\end{figure*}

\begin{figure*}[htp]
    \centering
    \includegraphics[width=1\linewidth]{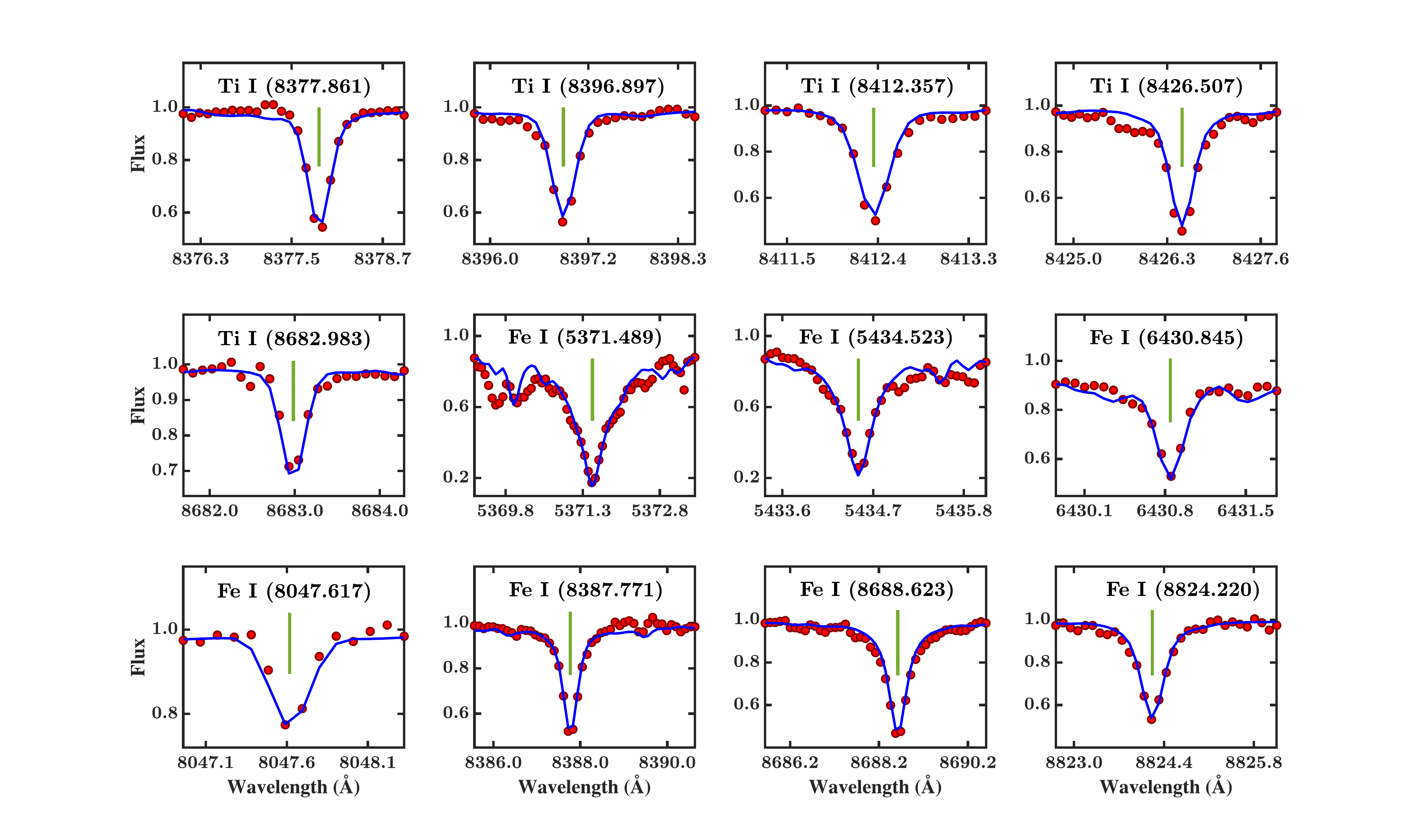}
    \caption{The continuation of Figure \ref{fig:bestfit_1} for the analyzed Ti I and Fe I lines. }
\label{fig:bestfit_2}
\end{figure*}

\begin{figure*}[htp]
    \centering
    \includegraphics[width=0.9\linewidth]{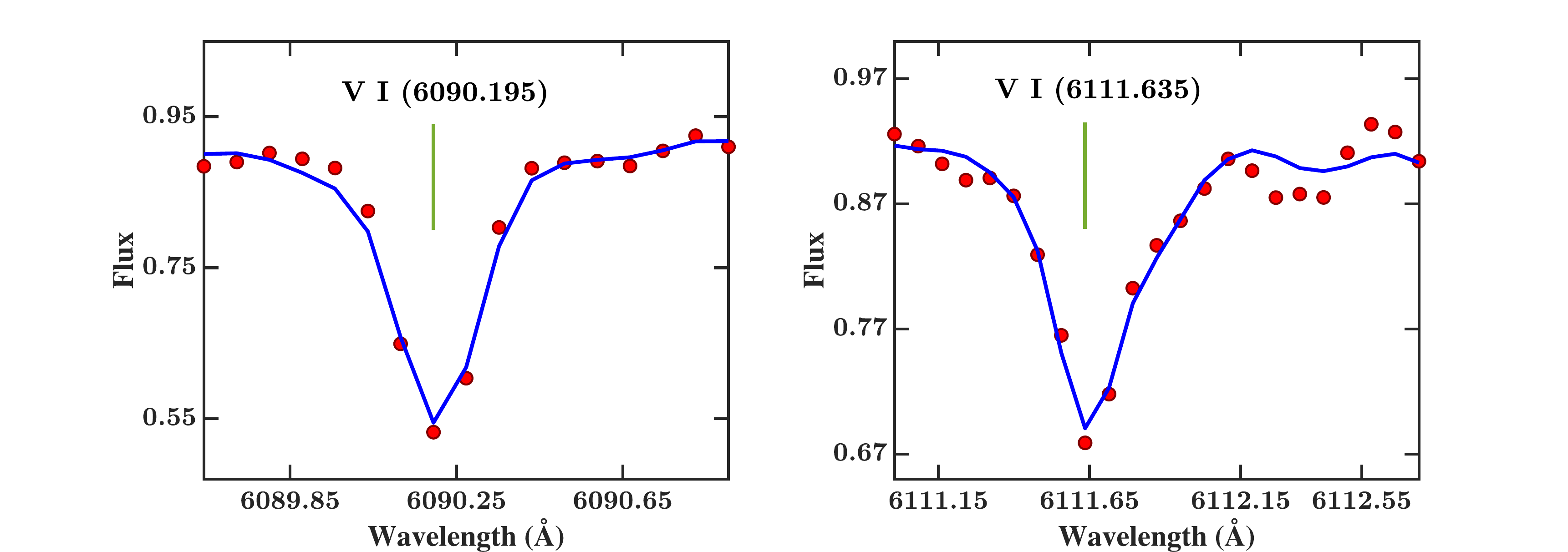}
    \caption{The continuation of Figures \ref{fig:bestfit_1} and \ref{fig:bestfit_2}  for the analyzed V I lines. The line data used in the spectral synthesis  includes the HFS splitting lines.}
\label{fig:bestfit_3}
\end{figure*}

\begin{figure}[htp]
    \centering
    \includegraphics[width=1\linewidth]{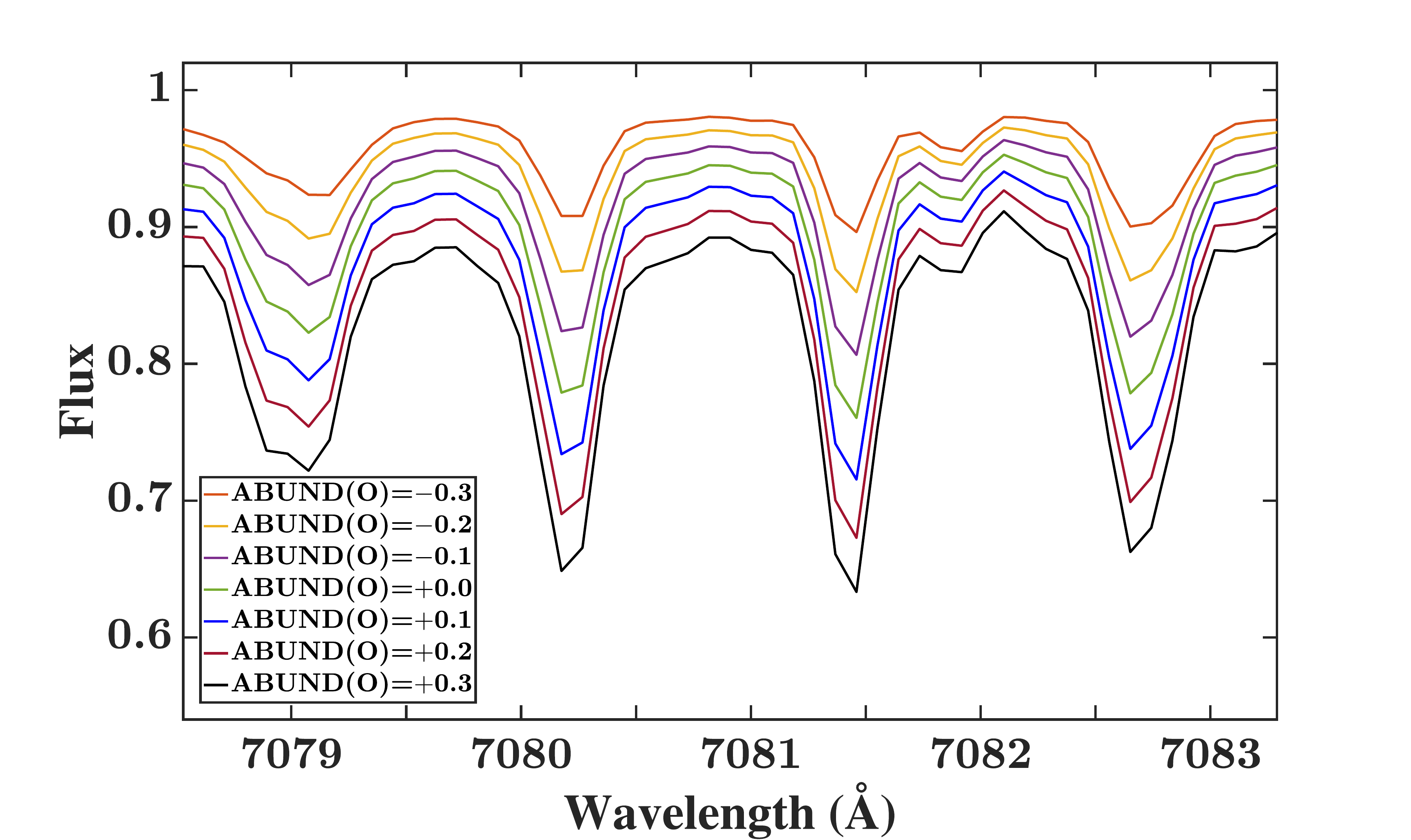}
    \caption{The continuum-normalized synthetic spectra associated with the star's parameters but with  different oxygen abundances while assuming ABUND(X)=0 for other elements X over a small portion of the $\gamma$ R$_{2}$ 0-0 TiO band}
\label{fig:O_variation_TiO_region}
\end{figure}

\begin{figure}[htp]
    \centering
    \includegraphics[width=1\linewidth]{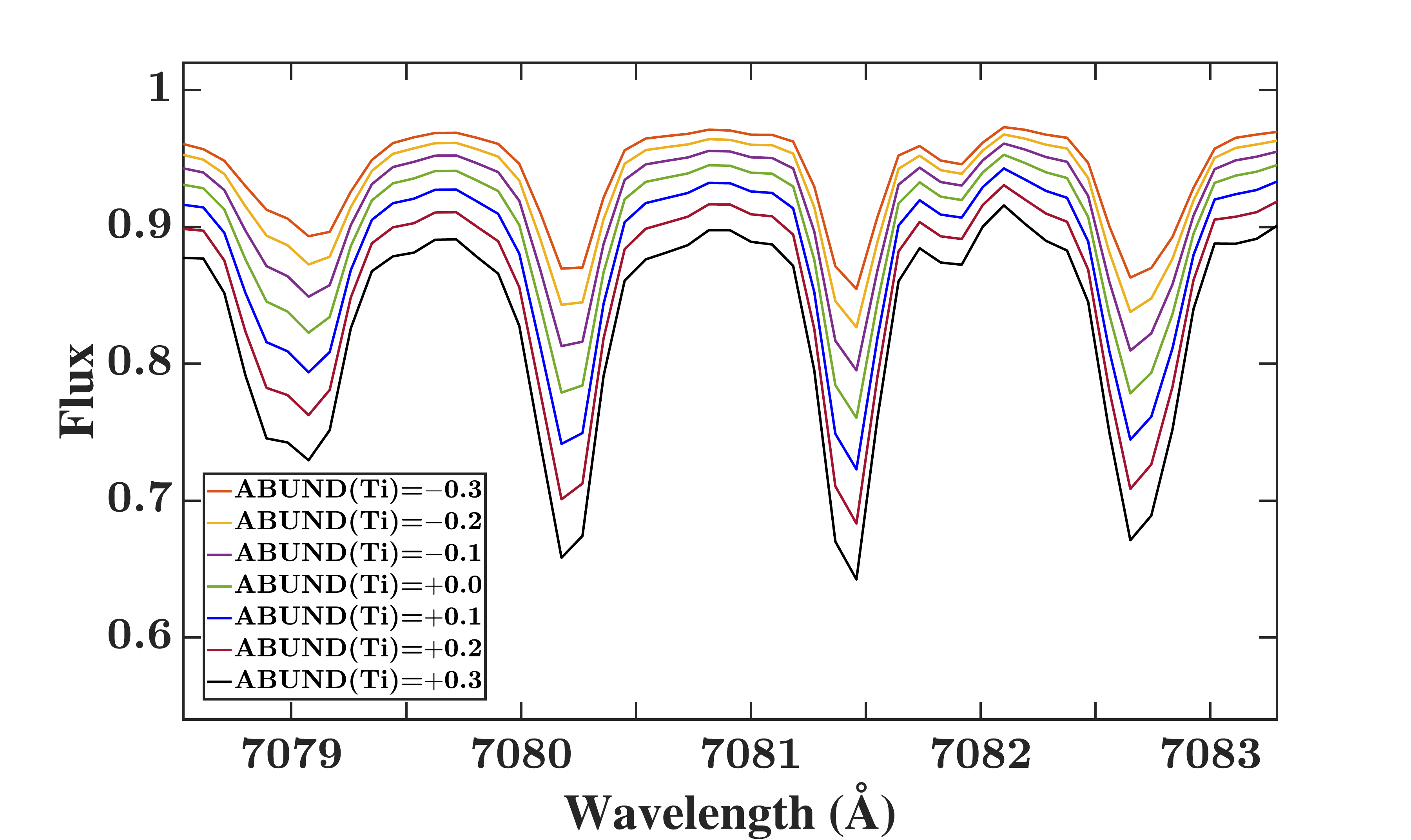}
    \caption{The continuum-normalized synthetic spectra associated with the star's parameters but with different titanium abundances while assuming ABUND(X)=0 for other elements X over a small portion of the $\gamma$ R$_{2}$ 0-0 TiO band}
\label{fig:Ti_variation_TiO_region}
\end{figure}

\begin{figure}[htp]
    \centering
    \includegraphics[width=1\linewidth]{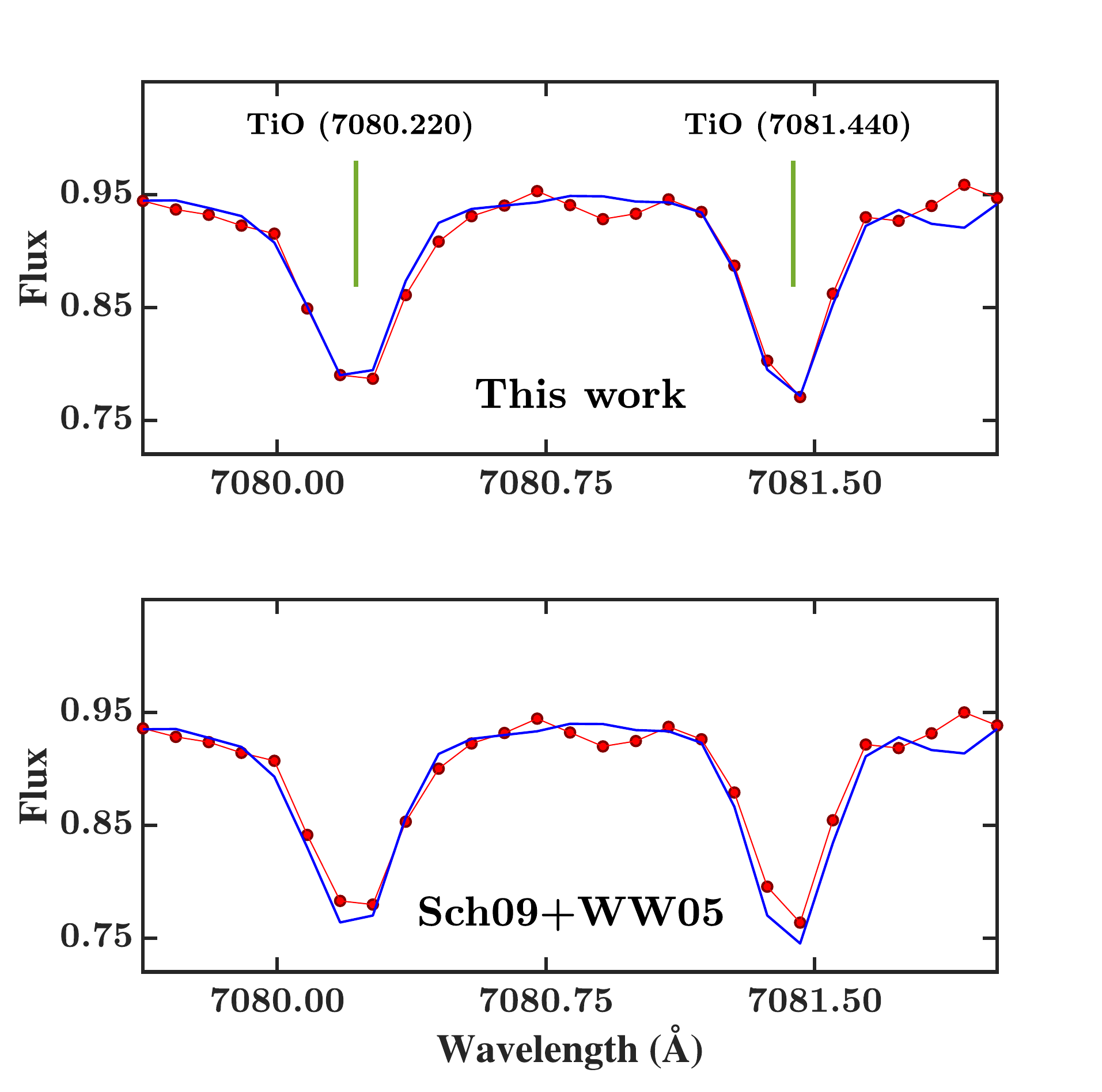}
    \caption{A small portion of the $\gamma$ R$_{2}$ 0-0 TiO band used in measuring the oxygen abundance. \textbf{Top:} Comparison between the normalized observed spectrum (red) and the best-fit synthetic model inferred from this study (blue).  \textbf{Bottom:} Comparison between the normalized observed spectrum (red) and the synthetic spectrum  associated with the star's physical parameters, Ca and V abundances from this study, but O, Ti and Fe abundances from previous studies, as reported in Table \ref{tab:results} (blue, Model$_{\rm WW05+Sch09}$). The observed spectrum is normalized to each corresponding synthetic model separately.}
\label{fig:TiO_comparison}
\end{figure}

\begin{figure}[htp]
    \centering
    \includegraphics[width=1\linewidth]{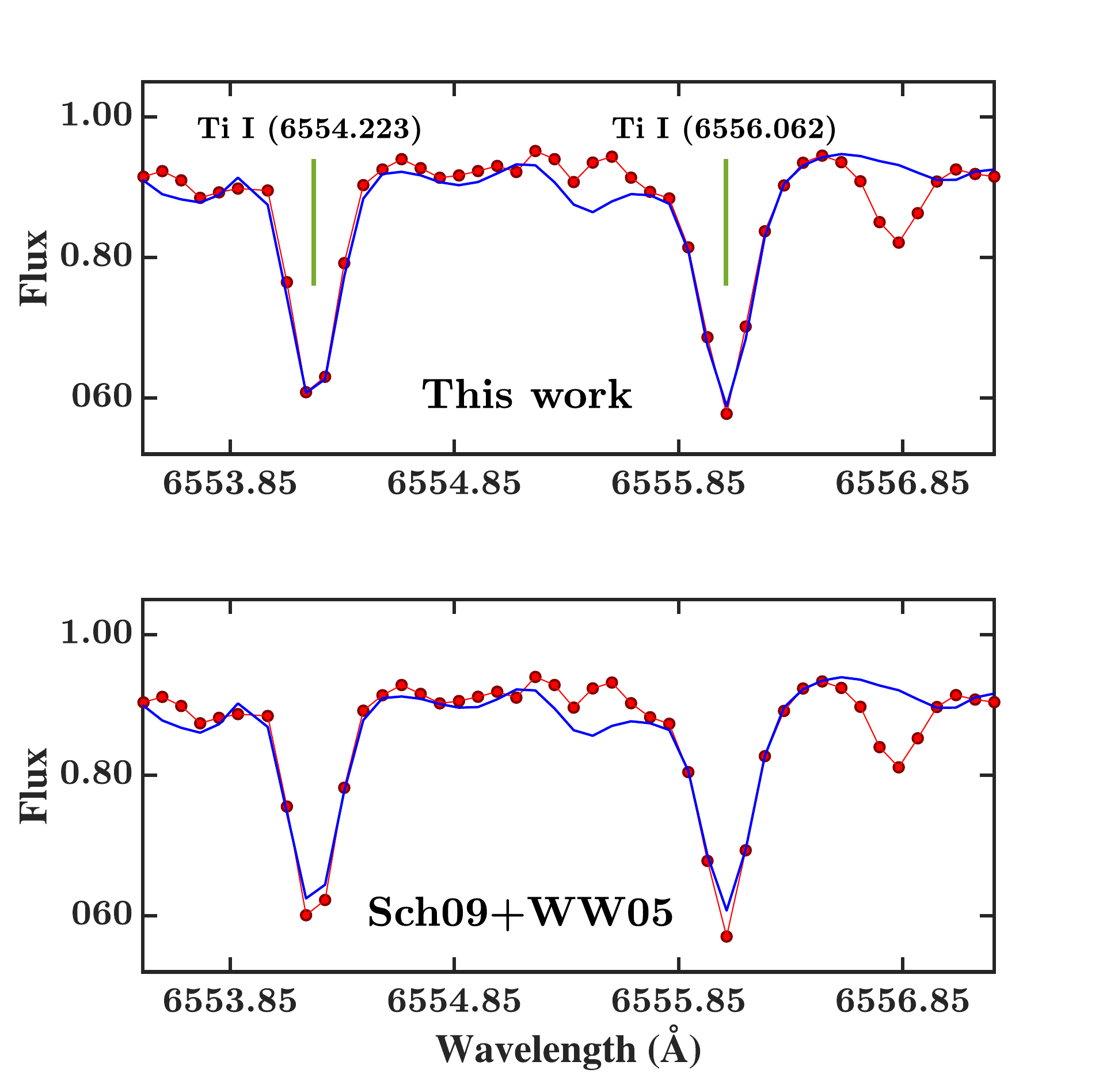}
    \caption{Identical to Figure \ref{fig:TiO_comparison}, but for two analyzed Ti I lines.}
\label{fig:Ti_comparison}
\end{figure}

\begin{figure}[htp]
    \centering
    \includegraphics[width=1\linewidth]{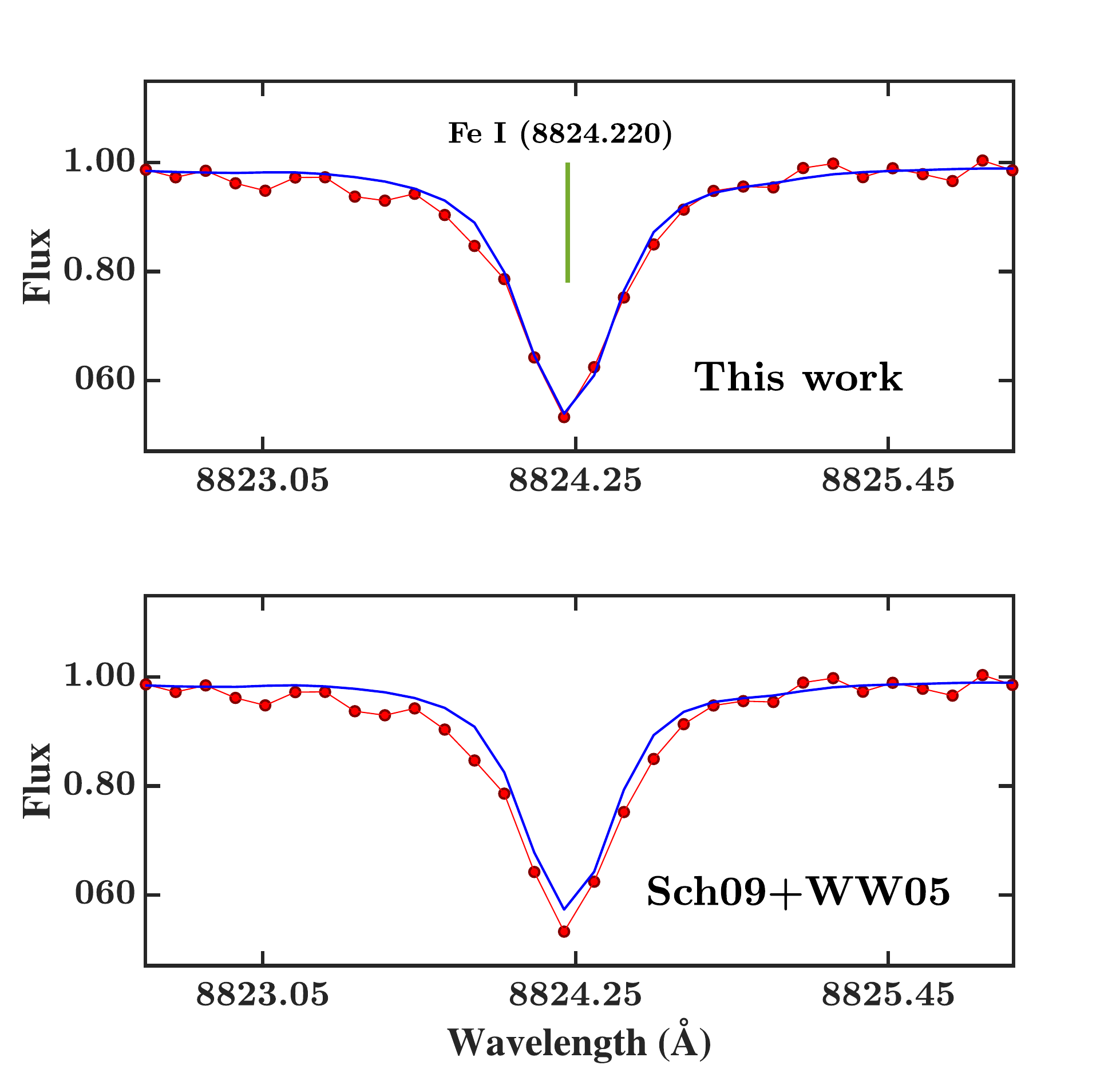}
    \caption{Identical to Figure \ref{fig:TiO_comparison}, but for one analyzed Fe I line.}
\label{fig:Fe_comparison}
\end{figure}

 Our results are shown in Table \ref{tab:results}: the chemical species (the first row), the number of analyzed lines relevant to each species (the second row), and the final derived abundances [X/H] (the third row). For comparison,  the elemental abundances from other studies, if any (the fourth row),  along with the respective references (the fifth row) are also demonstrated in this table. We compare the normalized observed spectrum (red dots) and the best-fit synthetic model  corresponding to the target's parameters and derived abundances (blue lines) for the analyzed atomic lines (but assuming ABUND(X)=0 for all other elements X that are not analyzed in this study) in Figures \ref{fig:bestfit_1}, \ref{fig:bestfit_2}, and \ref{fig:bestfit_3}. Clearly, there is a good agreement between the observed data and best-fit model for the majority of the spectral lines, which indicates the reliability of our technique.

\subsection{Abundance Errors}\label{sec:abundance_errors}
We estimate the uncertainty of the inferred abundances by the quadrature sum of the systematic and random (statistical) errors (Table \ref{tab:results}, rows 18 and 19, respectively). The random error is determined  using the standard  error of the mean, i.e., $\rm{\sigma_{ran}}$=std/$\rm{\sqrt{N}}$, where std is the standard deviation of  the abundances from different lines of each species. However, only two TiO lines (though both lines consist of a large number of TiO lines) and two V I lines (though both lines consist of multiple HFS lines) are analyzed (Table \ref{tab:line_data}), and random error does not apply to these elements. We note that the contribution of random errors is, in general, considerably smaller than that of systematic errors to  total uncertainties  (the last row of Table \ref{tab:results}, also see H23, H24, and H25).

The systematic errors are derived by determining the sensitivity of elemental abundances  to different physical stellar parameters. To this end, we deviate the four parameters, T$_{\rm eff}$, [M/H], log($g$), and $\xi$ by their associated uncertainties from WW05 (Table \ref{tab:properties}) in both negative (decreasing, hereafter, ``Neg'') and positive (increasing, hereafter, ``Pos'') directions one at a time (Table  \ref{tab:results}). We then run \texttt{AutoSpecFit} for each case (8 times) with only one deviated parameter in only one direction (Pos or Neg), while the other parameters are fixed to the target's parameters, and then measure the abundances of the analyzed elements. The abundance variations due to the perturbed parameters T$_{\rm eff}$, [M/H], log($g$), and $\xi$ are shown in rows 6-7, 9-10, 12-13 and 15-16 of Table \ref{tab:results}, respectively. The average of the absolute values of the two abundance variations due to the deviation of each parameter in the two Pos and Neg directions:

\begin{equation}\label{equ:average_abundance_variation}
\overline{\rm {\Delta}[X/H]}_{\rm P}=\frac{{\rm{\Delta}{\rm[X/H]}_{\rm{P,Neg}}}+{\rm{\Delta}{\rm[X/H]}_{\rm{P,Pos}}}}{2} \\   
\end{equation}

\noindent
where ``P'' denotes the parameters, which can be ``T'' for T$_{\rm eff}$, ``M'' for [M/H], ``G'' for log($g$), and ``$\xi$'' for $\xi$, are listed in rows 8, 11, 14, and 17 of Table \ref{tab:results}, respectively. The total systematic error $\rm{\sigma_{sys}}$ can be calculated by the quadrature sum of systematic errors due to the all four parameters P:

\begin{equation}\label{equ:sys_err}
{\rm \sigma_{sys}} = {\rm \sqrt{\sum_{{P}}^{} [\overline{\rm ({\Delta}[X/H])_{P}}]^2}} 
\end{equation}

\noindent
as shown in row 18 of Table \ref{tab:results}. The total abundance uncertainty is determined by the quadrature sum of the random (row 19)  and total systematic error:

\begin{equation}\label{equ:total_err}
\rm{\sigma_{tot}}=\sqrt{\sigma_{ran}^2 + \sigma_{sys}^2 }
\end{equation}

\noindent
as shown in the last row of Table \ref{tab:results}.

\section{Discussion}\label{sec:discussion}
Since  metal-poor stars are $\alpha$ enhanced, the abundances of O, Ca, and Ti, which are $\alpha$ elements, are markedly higher than the abundances of V and Fe for LHS 174 (Table \ref{tab:results}).  The abundances of different elements have different sensitivity to the variation of physical parameters. We discuss our results  and compare them to previous studies element-by-element in the following sections. In general, the differences in abundance measurement techniques, model atmospheres, and line lists used in different studies can give rise to discrepancies in the inferred  abundance vales of the same element.

\subsection{Oxygen Using TiO bands}
\cite{Valenti1998} first attempted to perform a detailed examination of TiO lines by modeling the $\gamma$ R$_{2}$ 0-0 TiO band (7078-7103 {\AA}) and fitting to the high-resolution (R$\sim$120000) spectrum of the M dwarf GJ 725B (M3.5V) observed using  the 2.7 m Harlan J. Smith telescope at McDonald Observatory. \cite{Bean2006} modified the method of \cite{Valenti1998} to fit these TiO bands using the high-resolution (R$\sim$50000) spectra of five M dwarfs  in binary systems with solar-type primaries (using the same telescope used by \cite{Valenti1998} but with a different instrument setup). Combined with the physical parameters and Fe and Ti abundances determined by WW05, Sch09 measured the abundance of oxygen for 62 late-K to early-M dwarfs, including LHS 174 using the same TiO fitting technique of \cite{Valenti1998} and \cite{Bean2006}. Figure \ref{fig:O_variation_TiO_region} shows the sensitivity of a portion of $\gamma$ R$_{2}$ 0-0 TiO band to the abundance of oxygen, which is similar to the sensitivity of this band to the abundance of titanium, as illustrated in Figure \ref{fig:Ti_variation_TiO_region}. The accurate measurement of oxygen abundance using TiO bands therefore requires a reliable abundance of titanium, which implies the necessity of the simultaneous abundance measurements of different elements during an iterative process.

Sch09 used only a small region of the above-mention TiO band from 7077 to 7084.5 {\AA} (excluding the band head) for their oxygen abundance measurements. However, as shown in Figure 1 of Sch09, merely a portion of this region, $\sim$7079.6-7082.0 {\AA}, was well modeled for LHS 174 using the updated version of model atmospheres from \cite{Hauschildt1999} and TiO line list from \citep{Plez1998}. We also find that this portion is properly modeled using our fitting method,  as shown in the top panel of Figure \ref{fig:TiO_comparison}, which can be used to infer the oxygen abundance of our target. Each of the two lines consists of a large number of TiO lines with a range of log($gf$) values as shown in Table \ref{tab:line_data}. The bottom panel of Figure \ref{fig:TiO_comparison} compares the normalized observed data with the synthetic model (hereafter, Model$_{\rm WW05+Sch09}$) generated using  our spectral synthesis method based on the star's physical parameters, Ca and V abundances from this study, but O, Ti, and Fe abundances from other studies, as reported in Table \ref{tab:results}. Although the abundances of O, Ti, and Fe from this study is consistent with those from WW05 and Sch09 within their uncertainties, our best-fit model clearly shows a better agreement with the observed spectrum, as compared to Model$_{\rm WW05+Sch09}$. 

As can be seen from Tables \ref{tab:results}, [O/H] is least sensitive to the variation of T$_{\rm eff}$, as compared to the other four elemental abundances. [O/H] has the second least sensitivity to [M/H] (with a minor difference) and log($g$) variations after [V/H]. However, oxygen abundance shows a moderate sensitivity to $\xi$. Due to its relatively low sensitivity to the physical parameters, [O/H] has the second least systematic error among the analyzed elements.

\subsection{Calcium}
Tables \ref{tab:results} and \ref{tab:results} show that [Ca/H] is  most sensitive to the variation of T$_{\rm eff}$ and log($g$), as compared to the other four elements under study, resulting in the largest systematic error, and in turn, the largest total uncertainty. The significantly high sensitivity to T$_{\rm eff}$ and log($g$) motivates us to employ the spectral lines of Ca as diagnostics for determining these two parameters in our future M-dwarf studies. On the other hand, calcium abundance is least sensitive to microturbulence $\xi$, while it has a moderate sensitivity to [M/H] relative to the other elements.

\subsection{Titanium}
As shown in Table \ref{tab:results}, relative to other analyzed elements, the abundance of Ti has a moderate sensitivity to the variation of the  two parameters  [M/H] and log($g$), but it is most sensitive to $\xi$. In addition, Titanium abundance has the second highest sensitivity to T$_{\rm eff}$ after [Ca/H]. While [Ti/H] from this work is consistent with that of from WW05 within their uncertainties, our Ti abundance shows a better agreement with the observed data. Similar to Figure \ref{fig:TiO_comparison}, Figure \ref{fig:Ti_comparison} compares the normalized observed spectrum with the best-fit model  from this work (the top panel)  and with  Model$_{\rm WW05+Sch09}$ (the bottom panel) for two Ti lines. Evidently, our best-fit model is better matched with the observed flux than Model$_{\rm WW05+Sch09}$.

It is important to note that WW05 measured the abundances of Ti and Fe using  the equivalent width (EQW) analysis. Although the EQW analysis is a robust technique to measure stellar elemental abundances, it can give rise to reliable abundances if the EQW of spectral lines is measured from the continuum level. This method is therefore mainly applicable to hotter FGK dwarfs with a well-defined continuum in their observed spectra. As shown in the top panel of Figure \ref{fig:TiO_comparison}, due to the significant contribution of prevailing  molecular lines, the fluxes of the (continuum-normalized) best-fit model in the regions around the Ti lines  are considerably below the unity, and for these lines, the continuum level cannot be identified when using the observed spectrum alone.  Synthetic spectral fitting presented in this study is thus a more reliable method to measure the elemental abundances of M dwarfs. 

\subsection{Vanadium}
\cite{Shan2021} characterized a number of V I lines within 800–910 nm wavelength region using high-resolution (R$~$94600) spectra from the CARMENES survey. They found that many of these lines had a distinctive broad and flat-bottom shape, resulting  from the HFS lines, which  needed to be incorporated in their atomic line list to infer a reliable vanadium abundance. However, the spectral resolution of our  target's spectrum is significantly lower than that of the CARMENES spectra, and we are unable to detect the flatness around the center of the two V I lines due to the HFS splitting lines. Nevertheless, we include these HFS lines (extracted from the VALD3 website) in our atomic line list  to generate more realistic synthetic spectra and determine a more accurate [V/H]. Our results in Tables \ref{tab:results} indicate that [V/H] has the least sensitivity to the variation of  [M/H] (almost along with [O/H]) and log($g$), and the second least sensitivity to T$_{\rm eff}$ after [O/H]. Vanadium abundance is moderately sensitive to $\xi$. The rather low sensitivity of [V/H] to the physical parameters has given rise to the smallest systematic error among the other studied elements.

\subsection{Iron}
As expected, our results in Tables \ref{tab:results} show that [Fe/H] has the highest sensitivity to the overall metallicity [M/H] relative to the other four elemental abundances. Historically, [Fe/H] is a proxy of the metallicity since Fe lines are abundant and rather easy to measure in stellar spectra. The Fe abundance is moderately sensitive to the deviation of the other three parameters T$_{\rm eff}$, log($g$) and $\xi$. Similar to Figures \ref{fig:TiO_comparison} and \ref{fig:Ti_comparison},  Figure \ref{fig:Fe_comparison} compares the normalized observed spectrum with the best-fit model from this work (the top panel) and with Model$_{\rm WW05+Sch09}$ (the bottom panel) for a single Fe line. Obviously, the observed data is more consistent with our best-fit model than Model$_{\rm WW05+Sch09}$. Analogous to [Ti/H], WW05 measured [Fe/H] using the EQW method. However, the flux level in spectral regions around some Fe I lines, such as those centered at 5371.489 {\AA}, 5434.523 {\AA}, and 6430.846 {\AA}, are clearly lower than the unity (Figure \ref{fig:bestfit_2}) , and the EQW analysis is not applicable to determine the abundances of these lines.

\section{Summary and Conclusion}\label{sec:summary}
We carry out an in-depth spectroscopic analysis of a metal-poor M subdwarf, LHS 174, using its high-resolution optical spectra. We identified 26 atomic lines appropriate for measuring the abundances of four elements Ca, Ti, V, and Fe as well as a small portion of a TiO band for inferring the abundance of O. As our pilot study, we use our technique, that originally developed for NIR spectra, to analyze the star's optical spectrum, indicating its applicability for measuring abundances of low-mass stars using different wavelength regimes. Our automatic pipeline, \texttt{AutoSpecFit} performs a normalization routine to normalize the observed flux relative to each single synthetic spectrum examined in each ${\chi}^{2}$ minimization process, which is of critical importance for a proper comparison between the star's spectrum and synthetic models. In addition,  the code takes into account the complex correlations between the abundances of different elements through an iterative procedure, where the abundances of all analyzed elements inferred from each iteration are incorporated into the  synthetic models generated for the next iteration. The process is repeated until globally consistent abundances of all studied elements  are achieved simultaneously.  Our results exhibit a good agreement between the observed spectrum and the best-fit model constructed based on the target's parameters and the elemental abundances from this work for the majority of spectral lines. We illustrated that the observed data is more matched with our best-fit model than the synthetic spectrum generated using the abundances from previous studies.

The accuracy of our inferred abundances is largely determined by the accuracy of the input physical parameters. Our primary motivation is thus to develop new techniques to reliably determine the physical parameters of M dwarfs.  For example,  the inferred parameter values can be improved by taking into account the degeneracy effect between different parameters. Our future follow-up research is an in-depth investigation of identifying spectral regions that are essentially sensitive to only one parameter. Utilizing the collected wavelength intervals will isolate the contribution of each parameter to the respective spectral lines and features and significantly reduce the degeneracy effect between parameters during spectral fitting.  The routine of modifying parameters along with abundances will be added to the future version of \texttt{AutoSpecFit}.  We emphasize that due to strong degeneracy between physical parameters and abundances, parameters should be varied and modified between iterations of abundance measurements, not together with abundances.

In addition, the accuracy of our analysis can be improved using spectra with higher SNR, which demands observations with larger telescopes for these low-mass, metal-poor stars. Higer-SNR spectra of metal-poor M subdwarfs would also allow us to measure the abundances of more elements, which could reveal essential information about the nucleosynthesis in the early times of the Galaxy's formation.

In addition to early-type, metal-poor M subdwarfs, our methodology is applicable to solar- and super-solar-metallicity  as well as cooler (down to $~$3100 K) M dwarfs, but using high-resolution, NIR spectra where H$_{2}$O molecular lines dominate (H24, H25, and our ongoing studies). In the NIR regions, there are a substantial number of lines that are minimally blended with the prevalent, background  H$_{2}$O lines, and can be used for abundance analysis. However, at lower temperatures ($\lesssim$3100 K), molecular lines become stronger, while the current molecular linelists are still insufficient to properly model many of these undesirable lines that significantly overlap with the lines of important species. This can also hinder to find appropriate normalizing intervals and perform the normalization procedure. Furthermore, MARCS models may not fully consider all physical conditions in the atmosphere of cooler, later type M dwarfs. For example, the dust cloud formation in stars with spectral types M6 or later may not be adequately treated in MARCS models. The problem with abundance measurements of more metal-rich and/or later-type M dwarfs  may be even more severe in high-resolution, optical spectra that include prominent TiO and other molecular lines. Nevertheless, there are still some optical wavelength ranges in early-type, metal-rich M dwarfs where the molecular lines are relatively weak and the spectral lines of some key elements can be identified, provided that the SNR is adequate. A new set of updated PHOENIX model atmospheres by Peter Hauschildt et al. (private communication) will become publicly available in the near future, which are expected to model the atmospheres of cool, late-type M dwarfs more accurately, and can be utilized in follow-up spectroscopic studies.

 \

\noindent
{Acknowledgments:}
 N.H. acknowledges support from NSF AAG grant No. 2108686 and from NASA ICAR grant No. NNH19ZDA001N. 
 T.N. acknowledges support from the Knut and Alice Wallenberg Foundation. 
 This work used the high-performance computing system of the physics and astronomy department at Georgia State University.

\clearpage

%\bibliography{Hejazi_etal_LHS174.bib}{}
%\bibliographystyle{aasjournal}

\end{document}